# Evaluating Classifiers in Detecting 419 Scams in Bilingual Cybercriminal Communities

Alex V Mbaziira
Dept of Computer Science
George Mason University
Fairfax, VA, USA
ambaziir@masonlive.gmu.edu

Ehab Abozinadah
Dept of Computer Science
George Mason University
Fairfax, VA, USA
eabozina@masonlive.gmu.edu

James H Jones Jr
Dept of Electrical Computer Engineering
George Mason University
Fairfax, VA, USA
jjonesu@gmu.edu

*Abstract*— **Incidents of organized cybercrime are rising because of criminals are reaping high financial rewards while incurring low costs to commit crime. As the digital landscape broadens to accommodate more internet-enabled devices and technologies like social media, more cybercriminals who are not native English speakers are invading cyberspace to cash in on quick exploits. In this paper we evaluate the performance of three machine learning classifiers in detecting 419 scams in a bilingual Nigerian cybercriminal community. We use three popular classifiers in text processing namely: Naïve Bayes, k-nearest neighbors (IBK) and Support Vector Machines (SVM). The preliminary results on a real world dataset reveal the SVM significantly outperforms Naïve Bayes and IBK at 95% confidence level.**

I. INTRODUCTION (HEADING 1)

Cybercrime has evolved from misuse and, or abuse of computer systems to sophisticated organized crime exploiting the internet. The causes of increasing incidents of cybercrime are attributed to: widespread internet access, increasing volume of internet-enabled devices and integration of social networking in computing architectures. These global internet-driven computing architectures continue to expand and build on top of existing immeasurable vulnerabilities, which provide miscreants with low barriers to commit and profit from cybercrime.

There are numerous types of cybercrime. Some research categorizes cybercrime into content-based and technology-based crime [1]. Other studies provide elaborate classification of cybercrime to include offences against confidentiality, availability and integrity of information and information technology [2]. In each category is a list of crimes that offer cybercriminals incentives and tools with capabilities to exploit computer system vulnerabilities for high financial rewards. Criminals also use the internet to obtain sophisticated tools for exploiting their victims without being detected or apprehended. Cyberspace provides criminals with capabilities for using dissociative anonymity to assume fake identifies for committing crime [3]. However, with social media, the true identities of cybercriminals can be leaked when the actor's friends in the criminal social network do not implement the same levels of privacy to hide their identities.

This study extends work in a previous paper [4] by implementing machine learning algorithms to detect 419 scams within an actual bilingual cybercriminal community. The main contribution of this paper is evaluation of the performance of machine learning algorithms in detecting 419 scams an actual bilingual cybercriminal community in a social network . We use in English as well as English and Nigerian Pidgin to evaluate the classifiers using the unigram and bigram models. We use three classifiers to detect 419 scammers within this cybercriminal community namely: Naïve Bayes, Support Vector Machines and k-Nearest Neighbor. Support Vector Machines significantly out-performed the other classifiers on datasets comprising of both English and Nigerian Pidgin unigram and bigram models at 95% confidence level. This because Nigerian Pidgin vocabulary has fewer words compared to English hence Support Vector  Machines tend to work well such datasets.

The rest of the paper is organized as follows: in Section 2 we discuss related work. In Section 3 we describe the dataset and criteria for evaluating the performance of these classifiers. In Section 4 we present the results and discussion of our experimental study and in section 5 we  draw our conclusions.

II. RELATED WORK

There is a growing body of research investigating the context and impact of cybercrime due to the increasing number of incidents and numerous vectors that criminals are exploiting to profit from crime [5], [6], [7], [8].  There are numerous types of cybercrime which are categorized as content-based and technology-based crime. Content-based cybercrime includes: scams, phishing, fraud, child pornography, spamming etc., while technology-based crime includes but is not limited to hacking, code injection, espionage [1].  In this section we review existing research on content-based crime in general but scams in particular. We also define scams and bilingual cybercriminal networks in context to this paper.

*A. Nigerian Bilingual Cybercriminals and 419 Scams*

This paper investigates  detection of 419 scams within a bilingual community of cybercriminals. The actors comprising the community of cybercriminals that we are studying was constructed into a graph in an earlier paper using publicly leaked emails obtained from an online data theft service [4]. These scams are known as advance-fee fraud or 419 scams [9], [10]. 419 scams originated from Nigerian in the 1970s at smaller scale but escalated in the 1980s during the oil boom as posted letters and then transitioned to email in the 1990's with commercialization of the Internet [11]. With time the origin of 419 scam cells expanded to different West African countries like Ghana, Cameroon, Ivory Coast, Benin as well other parts of the world. Although these scams usually go unreported, a 2013 report revealed that victims lost $12.7 billion during that year to this category of cyber-criminals [11].

Cybercriminals committing 419 scams speak at least two languages hence are bilingual. For purposes of this paper we use the term *bilingual cybercriminal community* to refer an online community of criminal actors that use English and Nigerian Pidgin to exploit victims using 419 scams. This because Nigeria as well as other West African countries with 419 scammers are very diverse countries with hundreds of local dialects. However, English and Nigerian Pidgin are the most popular and widely common spoken languages spoken in West Africa.

Nigerian Pidgin is an English-based pidgin comprising words from local Nigerian dialects and English. In Nigerian pidgin, the phrases are short compared to English while the English used in Nigerian pidgin does not follow proper grammar hence is broken English like any pidgin or Creole language.

*B. Content-based Cybercrime Detection*

Various research has studied detection of different types of content-based cybercrime like fraud, phishing and spam [12], [13]. Wang et al., study spam in social networks to build a social spam detection framework that filters spam across multiple social networks namely: MySpace, Twitter and WebbSpam Corpus [14]. Bosma et al., develop a social spam detection framework that uses link analysis and this implemented on a popular social network [15]. Bhat et al., propose a community-based framework and apply ensemble classifiers to detect spammers within community nodes in online social networks [16], [17]. Other studies evaluate predictive accuracy of several machine learning algorithms like Support Vector Machines, Random Forests, Naïve Bayes, Neural Networks in predicting phishing emails [18], [19]. Other research investigates the extent at which malware and spam has infiltrated online social networks [20]. However, these studies have not tackled bilingual datasets with 419 scams which are obtained from an actual cybercriminal community and evaluated performance of machine learning algorithms in detecting such scams within online cybercriminal communities. 419 scams comprise work-at-home scams, high yield investment scams, lottery scams or rewards from pay-per-click online adds.

*C. Machine Learning*

In this section we review supervised machine learning algorithms for our study. In supervised machine learning, the algorithms map inputs to specific outputs using input and output data [21]. We use three classifiers namely Naïve Bayes, Support Vector Machines and Decision Trees to detect scam in a social network of multi-lingual Nigerian cyber-criminals because these classifiers have been well studied and applied to spam and malware classification problems.

*a) Naïve Bayes:* this a popular classifier which has been applied to a variety of learning problems that are investigating scams like phishing, spamming and injected malicious hyperlinks. The algorithm implements Bayes Theorem which assumes conditional independence in feature variables of a learning set to predict statistical outcomes [22], [23].

*b) Support Vector Machines:* this another popular algorithm and that uses hyperplanes in dimensional space to address classification problems. This algorithm has been used in studying spam, fraud, malware, and phishing [24] [25].

k-Nearest Neighbors (kNN): this is also popular algorithm that uses instance-based learning to predict outcomes in learning problems. With instance-based learning, the kNN algorithm looks at the k-nearest neighbors when determining which instance to predict [26].

III. DATASET

We use a publicly leaked set of 1036 email addresses of Nigerian cybercriminals who are using an online data theft service called *PrivateRecovery* (which was formerly called *BestRecovery*) [27]. These cybercriminals are known for committing specific scams namely: advance-fee, online dating and *Nigerian letter* scams. Facebook lookups were conducted on each email address to identify corresponding public profiles of the criminal actors and their friends. The Facebook URLs of these actors was used in a previous paper to construct large graph of 43,125 criminal nodes [4]. These Facebook accounts of these criminal actors are real because the actors post and share a lot of personal information in form of text and photographs. The average number of friends for the 150 important criminal actors is 490 while the 4966 is the maximum number of friends these actors have. For this study, we used public data from 150 criminal nodes which had a high PageRank. During data collection, we did not engage with or friend the actors through their Facebook accounts.

*A. Dataset Description*

For our experimental study, we first generate two primary datasets from records which are randomly selected from the 150 Facebook accounts with high PageRank scores. Primary Dataset 1 (PD1) has English only records while Primary Dataset 2 (PD2) has half of the records in English and the other half in Nigerian Pidgin as shown in Table 1. The data in each primary dataset is labeled and then preprocessed to remove all non ASCII characters, symbols and punctuation marks except for the apostrophes, which we escaped. The data used in our classification problem is in two languages namely English and Nigerian Pidgin both of which use Latin characters hence do not use special symbols or non ASCII characters which is typical in languages like French, Spanish etc that use such characters to emphasize accents for certain words. However, in the data there was some evidence of use non ASCII characters in form on text-based emoticons expressing emotion. We do not stem the English words in the sub-datasets but use term frequency-inverse document frequency (tf-idf) to weight the words.

From each of the primary datasets we obtained two sub-datasets of unigram words and bigram words as shown in Table I. Sub-datasets (SD) A and B contains English unigram and bigram words respectively while Sub-datasets (SD) C and D has both unigram and bigram words respectively in both English and Nigerian Pidgin.

### B. Classifier Evaluation Metrics

Our study uses binary classification to train and test text instances in the datasets as either scam or not-scam. To evaluate our classifiers we use *Recall*, *Precision* and *F1 measure* on unigram and bigram word vectors. *Recall* measures the percentage of scam messages that are detected hence this metric determines how well a classifier performs in identifying a condition. *Precision,* however, measures how many of the scam messages are detected correctly hence this is a measure of probability that a predicted outcome is the right one [28]. F1 measure is a harmonic mean of precision and recall.

Let $x_{ns \to ns}$ be the number of *not-scam* posts classified as not-*scam*, $x_{ns \to s}$ be the number of not-scam posts misclassified as scam, $x_{s \to ns}$ be the number of *scam* posts misclassified as *not-scam* and $x_{s \to s}$ be the scam posts classified as scam. Therefore the equations for recall, precision and F1 will be:

$$Recall = \frac{x_{s \to s}}{x_{s \to s} + x_{s \to ns}} \quad (1)$$

$$Precision = \frac{x_{s \to s}}{x_{s \to s} + x_{ns \to s}} \quad (2)$$

$$F1\ Measure = \frac{2PR}{P+R} \quad (3)$$

### C. Experimental Setup

In this section we demonstrate how we obtain results on performance of the three classifiers in WEKA[29] using the four sub-datasets. To solve our classification problem we use three classifiers namely: Naïve Baye*s (NB)*[30], Support Vector Machines (*LibSVM*), and k-Nearest Neighbor (*IBK*) [26]. To obtain the results, each of the random sample sub-datasets is split into training and testing tests. 80% of the data in each sub-dataset is randomly allocated for training and 20% for testing. We also used 10-fold cross validation method to improve the performance of the classifiers. Using cross validation, each of the four sub-datasets were split up into 10 sets of equal proportion. Training was done on nine sets while testing is done of one. This process was repeated 10 times to ensure independence of the elements in the sample and also to minimize biases in the outcomes.

TABLE I. DESCRIPTION OF DATASETS USED IN ANALYSIS

| PD # | SD # | Language | N-Gram Words | # Words |
|---|---|---|---|---|
| 1 | A | English | Unigram | 2081 |
| 1 | B | English | Bigram | 12070 |
| 2 | C | English & Nigeria Pidgin | Unigram | 1875 |
| 2 | D | English & Nigeria Pidgin | Bigram | 3057 |

TABLE II. PRECISION, RECALL, F-MEASURE, ROC CURVE AREA, PRECISION-RECALL CURVE FOR ENGLISH UNIGRAM WORDS USING SUB-DATASET A

| Classifier | Precision | Recall | F-Measure | ROC Area | PRC |
|---|---|---|---|---|---|
| NB | 0.915 | 0.911 | 0.911 | 0.964 | 0.96 |
| LIBSVM | 0.886 | 0.885 | 0.885 | 0.947 | 0.945 |
| IBK | 0.833 | 0.78 | 0.771 | 0.822 | 0.811 |

TABLE III. PRECISION, RECALL, F-MEASURE, ROC CURVE AREA AND PRECISION-RECALL CURVE FOR ENGLISH BIGRAM WORDS USING SUB-DATASET B

| Classifier | Precision | Recall | F-Measure | ROC Area | PRC |
|---|---|---|---|---|---|
| NB | 0.72 | 0.565 | 0.473 | 0.895 | 0.883 |
| LIBSVM | 0.673 | 0.656 | 0.648 | 0.742 | 0.734 |
| IBK | 0.695 | 0.515 | 0.371 | 0.644 | 0.659 |

TABLE IV. PRECISION, RECALL, F-MEASURE, ROC CURVE AREA PRECISION-RECALL CURVE FOR ENGLISH AND NIGERIAN PIDGIN UNIGRAM WORDS USING SUB-DATASET C

| Classifier | Precision | Recall | F-Measure | ROC Area | PRC |
|---|---|---|---|---|---|
| NB | 0.964 | 0.964 | 0.964 | 0.994 | 0.994 |
| LIBSVM | 0.962 | 0.962 | 0.962 | 0.993 | 0.994 |
| iBK | 0.851 | 0.79 | 0.781 | 0.915 | 0.921 |

TABLE V. PRECISION, RECALL, F-MEASURE, ROC CURVE AREA AND PRECISION-RECALL CURVE FOR ENGLISH AND NIGERIAN PIDGIN BIGRAM WORDS USING SUB-DATASET D

| Classifier | Precision | Recall | F-Measure | ROC Area | PRC |
|---|---|---|---|---|---|
| NB | 0.887 | 0.861 | 0.859 | 0.981 | 0.981 |
| LIBSVM | 0.898 | 0.895 | 0.895 | 0.94 | 0.928 |
| iBK | 0.844 | 0.796 | 0.789 | 0.901 | 0.909 |

### IV. EXPERIMENTAL RESULTS

#### A. Results of Evaluation Metrics

In this section we first present the experimental results for performance of the classifiers on unigram and bigram words for the four sub-datasets. The evaluate the classifiers we use precision, recall, F-measure, ROC Curve and PR-Curve on datasets.

Table II shows the results for performance of the three classifiers using sub-dataset A of English unigrams. The results in this table reveal has precision of 0.915, recall of 0.911, f-measure of 0.911, ROC Area of 0.964 and PRC of 0.96. LibSVM has a precision of 0.866, recall of 0.885, f-measure of 0.885, ROC Area of 0.947 and PR-curve of 0.945. IBK has a precision of 0.833, recall of 0.78, f-measure of 0.771, ROC-curve of 0.822 and PR-curve of 0.811.

Table III presents results of the 3 classifiers using sub-dataset B of English bigram words. Detailed results in this table indicate that Naïve Bayes has a precision of 0.72. recall of 0.565, f-measure of 0.473, ROC Area of 0.895 and PR curve of 0.883. Comparatively, LibSVM has precision of 0.673, recall of 0.656, f-measure of 0.648, ROC Area of 0.742 and PR-Curve of 0.734. IBK has precision of 0.695, recall of 0.515, f-measure of 0.371, ROC of 0.644 and PR-curve of 0.659.

Table IV shows results for the classifier performance on sub-dataset C which contains unigram words in both English and Nigerian Pidgin. The results indicate that Naïve Bayes has a precision of 0.964, recall of 0.964, f-measure of 0.964, ROC area of 0.994 and PR-curve of 0.994. LibSVM has a precision of 0.962, recall of 0.962, f-measure of 0.962, ROC area of 0.963 and PR-curve of 0.994. IBK has a precision of 0.851, recall of 0.79, f-measure of 0.781, ROC area of 0.915 and PR-curve of 0.921.

Table V indicates results for performance of the three classifiers on sub-dataset D which contains bigrams words in both English and Nigerian Pidgin. The results in this table indicate that LibSVM has a precision of 0.898, recall of 0.895, f-measure of 0.895, ROC Area of 0.94 and PR curve of 0.928. Naïve Bayes has a precision of 0.887, recall of 0.861, f-measure of 0.859, ROC area of 0.981 and PR-curve of 0.981. Finally, IBK has precision of 0.844, recall of 0.796, f-measure of 0.789, ROC area of 0.901 and PR-curve of 0.909.

*B. Classifier Performance Evaluation*

In this section we evaluate performance of the classifiers to determine the best classifier for detecting scam within this community of bilingual cybercriminals using unigram and bigram models. We evaluate LibSVM against Naïve Bayes and IBK to establish the significance of results at 95% confidence level using the four datasets. To achieve this we use a 2-tailed T-test evaluate performance metrics of LibSVM against Naïve Bayes and IBK on the four sub-datasets. To perform this test, we run the experiment five times and for each run we perform 10-fold cross validation. During each run the instances are randomized and the dataset is split into 80% training test and 20% testing set.

The performance metrics that we use to evaluate performance of our classifiers on the sub-datasets are ROC area, PR-curve and f-measure. We develop several hypotheses to test significance of the outcomes of the classifiers predicting accuracy in detecting 419 scams on datasets with English only as well as English and Nigerian Pidgin using unigram and bigram models. We use $H_0$ to represent the null hypothesis and $H_1$ to represent the alternate hypothesis. We compare the performance of LibSVM with Naïve Bayes and IBK on the four sub-datasets.

TABLE VI. EVALUATING *LIBSVM* AGAINST OTHER CLASSIFIERS USING *ROC CURVE AREA* AT 95% CONFIDENCE (± FOR STANDARD DEVIATION)

| SD # | LibSVM | LibSVM vs NB | Hypothesis ($\alpha$=0.05) | LibSVM vs IBK | Hypothesis ($\alpha$=0.05) |
|---|---|---|---|---|---|
| A | 0.93±0.02 | 0.95 ±0.01 | Not Reject | 0.80 ±0.02 | Reject |
| B | 0.94±0.03 | 0.88 ±0.03 | Not Reject | 0.84 ±0.12 | Not Reject |
| C | 0.99±0.00 | 1.00 ±0.00 | Accept | 0.92 ±0.02 | Reject |
| D | 0.89±0.02 | 0.99 ±0.00 | Accept | 0.94 ±0.02 | Accept |

We evaluate classifier performance using ROC area as below:

- $H_0$: LibSVM's ROC area is greater than IBK's ROC Area for English unigrams while for $H_1$: LibSVM's ROC area is not greater that IBK's ROC Area for English unigrams. We reject the null hypothesis $H_0$ because LibSVM's ROC area is significantly worse at 0.8 with a standard deviation of 0.02 as shown in Table VI.
- $H_0$: LibSVM's ROC area is greater than Naïve Bayes' ROC Area for both English and Nigerian Pidgin unigrams while for $H_1$: LibSVM's ROC area is not greater that Naïve Bayes ROC Area for English and Nigerian Pidgin unigrams. We accept the null hypothesis $H_0$ because LibSVM's ROC area is significantly better at 1.00 as shown in Table VI.
- $H_0$: LibSVM's ROC area is greater than IBK's ROC Area for English and Nigerian Pidgin unigrams while for $H_1$: LibSVM's ROC area is not greater that IBK's ROC Area for English and Nigerian Pidgin unigrams. We reject the null hypothesis $H_0$ because LibSVM's ROC area for both English and Nigerian unigrams is significantly worse at 0.92 and standard deviation of 0.02 as shown in Table VI.
- $H_0$: LibSVM's ROC area is greater than Naïve Bayes' ROC area for English and Nigerian Pidgin bigrams while for $H_1$: LibSVM's ROC area is not greater that Naïve Bayes' ROC area for English and Nigerian Pidgin bigrams. We accept the null hypothesis $H_0$ because LibSVM's ROC area for both English and Nigerian bigrams is significantly better at 0.99 as shown in Table VI.

$H_0$: LibSVM's ROC area is greater than IBK's ROC area for English and Nigerian Pidgin bigrams while for $H_1$: LibSVM's ROC area is not greater that IBK's ROC area for English and Nigerian Pidgin bigrams. We accept the null hypothesis $H_0$ because LibSVM's ROC area for both English and Nigerian bigrams is significantly better at 0.94 and standard deviation of 0.02 as shown in Table VI.

Here we continue the evaluation for classifier performance using PR area as shown below:

- $H_0$: LibSVM's PR area is greater than IBK's PR area for English unigrams while for $H_1$: LibSVM's PR area is not greater than IBK's PR area for English unigrams. We reject the null hypothesis $H_0$ because LibSVM's PR area for English unigrams is significantly worse at 0.79 and standard deviation of 0.02 as shown in Table VII.
- $H_0$: LibSVM's PR area is greater than Naïve Bayes's PR area for English bigrams while for $H_1$: LibSVM's PR area is not greater than Naïve Bayes' PR area for English bigrams. We reject the null hypothesis $H_0$ because LibSVM's PR area for English bigrams is significantly worse at 0.86 and standard deviation of 0.02 as shown in Table VII.

- $H_0$: LibSVM's PR area is greater than Naïve Bayes's PR area for English and Nigerian Pidgin unigrams while for $H_1$ : LibSVM's PR area is not greater than Naïve Bayes' PR area for English and Nigerian Pidgin unigrams. We accept the null hypothesis $H_0$ because LibSVM's PR area for English and Nigerian Pidgin unigrams is significantly better at 1.00 as shown in Table VII.
- $H_0$: LibSVM's PR area is greater than IBK's for English and Nigerian Pidgin unigrams while for $H_1$ : LibSVM's PR area is not greater than IBK's PR area for English and Nigerian Pidgin's unigrams. We reject the null hypothesis $H_0$ because LibSVM's PR area for English and Nigerian Pidgin unigrams is significantly worse at 0.93 and standard deviation of 0.01 as shown in Table VII.
- $H_0$: LibSVM's PR area is greater than Naïve Bayes's PR area for English and Nigerian Pidgin bigrams while for $H_1$ : LibSVM's PR area is not greater than Naïve Bayes' PR area for English and Nigerian Pidgin bigrams. We accept the null hypothesis $H_0$ because LibSVM's PR area for English and Nigerian Pidgin bigrams is significantly better at 0.99 as shown in Table VII.
- $H_0$: LibSVM's PR area is greater than IBK's PR area for English and Nigerian Pidgin bigrams while for $H_1$ : LibSVM's PR area is not greater than IBK's PR area for English and Nigerian Pidgin bigrams. We accept the null hypothesis $H_0$ because LibSVM's PR area for English and Nigerian Pidgin bigrams is significantly better at 0.94 and standard deviation of 0.02 as shown in Table VII.

TABLE VII. EVALUATING *LIBSVM* AGAINST OTHER CLASSIFIERS USING *PR CURVE AREA* AT 95% CONFIDENCE (± FOR STANDARD DEVIATION)

| SD # | LibSVM | LibSVM vs NB | Hypothesis (α=0.05) | LibSVM vs IBK | Hypothesis (α=0.05) |
|---|---|---|---|---|---|
| A | 0.93±0.02 | 0.95 ±0.01 | Not Reject | 0.79 ±0.02 | Reject |
| B | 0.94±.02 | 0.86 ±0.02 | Reject | 0.82 ±0.10 | Not Reject |
| C | 0.99±0.00 | 1.00 ±0.00 | Accept | 0.93 ±0.01 | Reject |
| D | 0.89±0.02 | 0.99 ±0.00 | Accept | 0.94 ±0.02 | Accept |

TABLE VIII. EVALUATING *LIBSVM* AGAINST OTHER CLASSIFIERS USING *F-MEASURE* AT 95% CONFIDENCE (± FOR STANDARD DEVIATION)

| SD # | LibSVM | LibSVM vs NB | Hypothesis (α=0.05) | LibSVM vs IBK | Hypothesis (α=0.05) |
|---|---|---|---|---|---|
| A | 0.86±0.03 | 0.89 ±0.02 | Not Reject | 0.76 ±0.01 | Reject |
| B | 0.80±0.05 | 0.48 ±0.04 | Reject | 0.37 ±0.02 | Reject |
| C | 0.94±0.01 | 0.97 ±0.01 | Accept | 0.79 ±0.03 | Reject |
| D | 0.77±0.04 | 0.85 ±0.03 | Accept | 0.79 ±0.04 | Not Reject |

We conclude the evaluation for classifier performance with f-measure as below:

- $H_0$: LibSVM's f-measure is greater than IBK's f-measure for English unigrams while for $H_1$ : LibSVM's f-measure is not greater than IBK's f-measure for English unigrams. We reject the null hypothesis $H_0$ because LibSVM's f-measure for English unigrams is significantly worse at 0.76 and standard deviation of 0.01 as shown in Table VIII.
- $H_0$: LibSVM's f-measure is greater than Naïve Bayes' f-measure for English bigrams while for $H_1$ : LibSVM's f-measure is not greater than Naïve Bayes' f-measure for English bigrams. We reject the null hypothesis $H_0$ because LibSVM's f-measure for English bigrams is significantly worse at 0.48 and standard deviation of 0.04 as shown in Table VIII.
- $H_0$: LibSVM's f-measure is greater than IBK's f-measure for English bigrams while for $H_1$ : LibSVM's f-measure is not greater than IBK's f-measure for English bigrams. We reject the null hypothesis $H_0$ because LibSVM's f-measure for English bigrams is significantly worse at 0.37 and standard deviation of 0.02 as shown in Table VIII.
- $H_0$: LibSVM's f-measure is greater than Naïve Bayes' f-measure for English and Nigerian Pidgin unigrams while for $H_1$ : LibSVM's f-measure is not greater than Naïve Bayes' f-measure for English and Nigerian Pidgin unigrams. We accept the null hypothesis $H_0$ because LibSVM's f-measure for English and Nigerian Pidgin unigrams is significantly better at 0.97 and standard deviation of 0.01 as shown in Table VIII.
- $H_0$: LibSVM's f-measure is greater than IBK's f-measure for English and Nigerian Pidgin unigrams while for $H_1$ : LibSVM's f-measure is not greater than IBK's f-measure for English and Nigerian Pidgin unigrams. We reject the null hypothesis $H_0$ because LibSVM's f-measure for English and Nigerian Pidgin unigrams is significantly worse at 0.79 and standard deviation of 0.03 as shown in Table VIII.
- $H_0$: LibSVM's f-measure is greater than Naïve Bayes f-measure for English and Nigerian Pidgin bigrams while for $H_1$ : LibSVM's f-measure is not greater than Naïve Bayes' f-measure for English and Nigerian Pidgin bigrams. We accept the null hypothesis $H_0$ because LibSVM's f-measure for English and Nigerian Pidgin bigrams is significantly better at 0.85 and standard deviation of 0.03 as shown in Table VIII.

As shown in Tables VI, VII and VII, 8 of the null hypotheses are accepted while 9 hypotheses are rejected and 6 hypotheses are not rejected. All the 8 null hypotheses which are accepted reveal that LibSVM significantly outperformed other classifiers on a unigram and bigram models that comprise both English and Nigerian Pidgin words. The rejected hypotheses reveal that IBK performed significantly worse compared to LibSVM mainly on the English only unigram and bigram models as well as on the unigram model comprising Nigerian Pidgin and English words. The 6 hypotheses that are not rejected were based on unigram and bigram model for English only words.

The LibSVM out-performed other classifiers on English and Nigerian Pidgin unigram and bigram model because these sub-datasets had fewer words in their vocabulary compared to the English words. This is because Nigerian Pidgin uses a limited vocabulary of words which are selected from both English and other local Nigerian dialects

## V. Conclusion

This study evaluated performance of three classifiers in detecting 419 scams within a bilingual cybercriminal community. The three classifiers we used are LibSVM, Naïve Bayes and IBK. We evaluated the performance of the three classifiers using both unigram and bigram models comprising and of English words as well as both English and Nigerian Pidin words. In both models, LibSVM outperformed Naïve Bayes and IBK. We used a 2-tailed t-test at 95% confidence to evaluate the classifiers on both the unigram and bigram models of English words as well as both English and Nigerian Pidgin words. These results motivate future work to explore the use of ensemble learning in detecting scams in bilingual criminal communities.

## Acknowledgment

We would like to thank Prof Damon McCoy for his initial input and allowing us to use the publicly leaked emails from which we collected data used in the earlier paper. We also want to thank all the anonymous reviewers whose comments were used to improve this paper.